\documentclass[11pt]{article}
\usepackage{amssymb}
\usepackage{amsmath}
\usepackage{graphics}
\usepackage{epsfig}
\usepackage{a4wide}
\usepackage{slashed,cite}
\usepackage{color}
\usepackage{enumerate}
\usepackage{multirow}
\usepackage{booktabs}
\usepackage{diagbox}

\begin{document}

\newcommand{\nn}{\nonumber}
\newcommand{\ppHZ}{$pp \to HZ+X~$}
\newcommand{\ppqqHZ}{$pp \to q \bar{q} \to HZ+X~$}

\title{ $HZ$ associated production at the LHC within the littlest Higgs model at NLO+NLL accuracy }
\author{ Chen Liang-Wen, Ma Wen-Gan, Zhang Ren-You, Li Xiao-Zhou, and Wang Yong  \\
{\small  Department of Modern Physics, University of Science and Technology of China (USTC),}  \\
{\small   Hefei, Anhui 230026, People's Republic of China}
}
\date{}
\maketitle \vskip 15mm

\begin{abstract}
We consider the $HZ$ associated production at the $14~{\rm TeV}$ LHC in the littlest Higgs model (LHM) and study the corrections of the transverse momentum resummation and threshold resummation at the next-to-leading logarithmic (NLL) accuracy and the fixed-order prediction at the QCD next-to-leading order (NLO) including the contribution from the one-loop-induced $gg$-fusion channel. The QCD NLO+NLL effects on the integrated cross section and the distributions of transverse momentum and invariant mass of the $HZ$ system for the $HZ$ production in the LHM are discussed. The distributions of transverse momentum and invariant mass of the $HZ$ system are evaluated by means of the transverse momentum resummation and threshold resummation, respectively. We estimate their scale uncertainties and find that the predictions obtained at the QCD NLO+NLL accuracy are much more reliable than those using the pure NLO approach. We see also that the relative deviation between the results in the LHM and the standard model is considerably reduced by the resummation effects, but still observable.
\end{abstract}

%\fi

\vfill \eject \baselineskip=0.32in

\renewcommand{\theequation}{\arabic{section}.\arabic{equation}}
\renewcommand{\thesection}{\Roman{section}.}
\newcommand{\nb}{\nonumber}

%slash:
\newcommand{\Dir}{\kern -6.4pt\Big{/}}%su lettere italiane minuscole
\newcommand{\Dirin}{\kern -10.4pt\Big{/}\kern 4.4pt}
\newcommand{\DDir}{\kern -7.6pt\Big{/}}%su lettere italiane maiuscole
\newcommand{\DGir}{\kern -6.0pt\Big{/}}%su lettere greche

\makeatletter      % '@' is now a normal "letter" for TeX
\@addtoreset{equation}{section}
\makeatother       % '@' is restored as a "non-letter" character for TeX

\par
\section{INTRODUCTION}

\par
Since the Higgs boson discovery at the CERN Large Hadron Collider (LHC) in 2012\cite{Aad:2012tfa,Chatrchyan:2012ufa}, establishing the properties of the Higgs boson, especially its couplings to the standard model (SM) particles, has been one of the primary missions of the current LHC run. Furthermore, the so-called naturalness problem is still a haunting nightmare and is the driving force for new physics beyond the SM.

\par
The littlest Higgs model (LHM) is a prominent realization of the little Higgs mechanism, which is proposed to ameliorate the fine-tuning problem \cite{ArkaniHamed:2002qy,ArkaniHamed:2001nc,ArkaniHamed:2002qx}. In the LHM, a global $SU(5)$ symmetry and a locally gauged subgroup $G_{1} \otimes G_{2}=[SU(2)_{1} \otimes U(1)_{1}] \otimes [SU(2)_{2} \otimes U(1)_{2}]$ are introduced. The global symmetry $SU(5)$ is broken into its subgroup $SO(5)$ at the scale $f$. In the meantime, the local gauge symmetry $[SU(2) \otimes U(1)]^{2}$ is broken into its diagonal subgroup $SU(2)_{L}\otimes U(1)_{Y}$ spontaneously, which is identified as the SM electroweak gauge group. It is well known that the SM gauge bosons and the top quark contribute quadratic divergent terms to the Higgs boson mass. In the LHM, several heavy gauge bosons ($W_{H}^{\pm}$, $Z_{H}$, and $A_{H}$) and one heavy vectorlike quark ($T$) are introduced to cancel these quadratic divergences at the one-loop level. These additional heavy particles might exhibit signatures at the LHC.

\par
The associated $HZ$ production is one of the most important Higgs production channels at hadron colliders, and it is a direct process to investigate the $HZZ$ coupling. There have already been thorough efforts for precise predictions of the $pp \to HZ+X$ process. The next-to-leading-order (NLO) QCD and electroweak (EW) corrections have been calculated in Refs.\cite{Han:1991ia,Kniehl:1990iva,Ciccolini:2003jy}. The next-to-next-to-leading order (NNLO) QCD corrections also have been performed in Refs.\cite{Brein:2003wg,Brein:2012ne}.

\par
However, the fixed-order calculation is reliable only when all the scales are of the same order of magnitude. At the phase space boundaries, for example, when the $HZ$ system is produced with small $p_{T,HZ}$ or with invariant mass approaching the partonic center-of-mass energy, i.e., $z=M^2_{HZ}/\hat{s}\sim 1$, the coefficients of the perturbative expansion in $\alpha_s$ are enhanced by powers of large logarithms $\ln^m(M^2_{HZ}/p_{T,HZ}^2)$ or $\ln^m(1-z)/(1-z)$, which spoil the convergence of the fixed-order predictions. In order to obtain reliable results at the boundaries of the phase space, these large logarithms need to be resummed. The transverse momentum resummation technique \cite{Bozzi:2005wk,Catani:2000vq,Collins:1984kg} is proposed for the summation of the large logarithms of the type $\ln^m(M^2_{HZ}/p_{T,HZ}^2)$, and the threshold resummation technique \cite{Sterman:1986aj,Catani:1996yz,Catani:1989ne} for the summation of the large logarithms of the type $\ln^m(1-z)/(1-z)$.

\par
The transverse momentum resummation and the threshold resummation effects for $HZ$ production at the LHC in the SM were presented in Refs.\cite{Dawson:2012gs,Shao:2013uba}. The calculation for the NNLO QCD corrections to the SM Higgs boson production in association with a $Z$-boson at hadron colliders has been implemented by O. Brein {\it et al. } \cite{Brein:2003wg}. They find that the contribution from the lowest order $gg$-fusion channel at the LHC is more important than the other QCD NNLO corrections to $HZ$ production. The QCD NLO calculation of the $HZ$ production at the LHC within the framework of the LHM was provided in our previous work \cite{Zhang:2012vq}, where the effects of the LHM up to the QCD NLO from the $q\bar{q}$ annihilation channel were investigated, but the contribution from the $gg$-fusion channel was absent.

\par
In this work, we study the effects of the littlest Higgs model on the $HZ$ production at the QCD NLO and the next-to-leading-logarithmic (NLL) level including the lowest contribution from the $gg$-fusion channel. We organize the paper as follows. In Sec.II, we give a glance at the LHM theory. In Sec.III, we briefly describe the leading order (LO) and the QCD NLO calculation strategy, and recapitulate the well-known formalism of the transverse momentum resummation and the threshold resummation. The numerical analyses and discussions are presented in Sec.IV, where some numerical results of the integrated cross section and differential cross section by adopting the transverse momentum resummation and the threshold resummation are provided. Finally, a short summary is given. The related Feynman rules for the coupling vertices in the LHM are collected in the appendix.

\par
\section{BRIEF REVIEW OF THE LHM }
\par
The LHM is based on an $SU(5)/SO(5)$ nonlinear sigma model. The vacuum expectation value (VEV) breaks the global $SU(5)$ symmetry into its subgroup $SO(5)$ and at the same time breaks the local gauge symmetry
$[SU(2)_1\otimes U(1)_1]\otimes [SU(2)_2\otimes U(1)_2]$ into its diagonal subgroup $SU(2)_L\otimes
U(1)_Y$, which is identified as the electroweak gauge group in the SM. The gauge fields $W^{\prime\mu}$ and
$B^{\prime\mu}$ associated with the broken local gauge symmetries and the SM gauge fields can be expressed as follows:
\begin{equation}
W^{\mu}=sW_{1}^{\mu}+cW_{2}^{\mu}  \,,
~~~~~~
W^{\prime\mu}=-cW_{1}^{\mu}+sW_{2}^{\mu} \,,
\end{equation}
\begin{equation}
B^{\mu}=s^{\prime}B_{1}^{\mu}+c^{\prime}B_{2}^{\mu} \,,
~~~~~~~
B^{\prime \mu}=-c^{\prime}B_{1}^{\mu}+s^{\prime}B_{2}^{\mu} \,,
\end{equation}
where $s = \sqrt{1-c^2}$, $s^{\prime} = \sqrt{1-c^{\prime2}}$, and $c, c^{\prime}$ are given by
\begin{equation}
c=\frac{g_{1}}{\sqrt{g_{1}^{2}+g_{2}^{2}}} \,,\hspace{2.0cm}
c^{\prime}=\frac{g_{1}^{\prime}}{\sqrt{g_{1}^{\prime2}+g_{2}^{\prime2}}} \,.
\end{equation}

\par
At the scale $f$, the SM gauge bosons remain massless, while the heavy gauge bosons acquire masses of order $f$. The $W$ and $B$ are identified as the SM gauge bosons, with couplings of $g=g_1 s=g_2 c$ and $g^{\prime}=g_1^{\prime} s^{\prime}=g_2^{\prime} c^{\prime}$. The electroweak symmetry breaking gives the masses for the SM gauge bosons and induces further mixing between the light and heavy gauge bosons. We denote the light gauge boson mass eigenstates as $W^{\pm}_{L}$, $Z_{L}$, and $A_{L}$ (i.e., $W^{\pm}$, $Z$, and $\gamma$)
and the new heavy gauge boson mass eigenstates as $W_{H}^{\pm}$, $Z_{H}$, and $A_{H}$. The masses of these gauge bosons to the order of $v^2/f^2$ are given by \cite{lhest1}
\begin{eqnarray}
&&
M_{W^{\pm}}^2=M_{W_L^{\pm}}^2 = m_W^2 \left[1 - \frac{v^2}{f^2}
\left( \frac{1}{6}+ \frac{1}{4} (c^2-s^2)^2 \right) + 4
\frac{v^{\prime 2 }}{v^2}\right] \,, \nonumber \\
&&
M_{Z}^{2}=M_{Z_{L}}^{2}=m_{Z}^{2}\left\{1-\frac{v^{2}}{f^{2}}\left[\frac{1}{6}
+\frac{1}{4}(c^{2}-s^{2})^{2}+
\frac{5}{4}(c^{\prime2}-s^{\prime2})^{2}-\frac{\chi^{2}}{2}\right]\right\} \,, \\
&&
M_{\gamma}^{2}=0 \,, \nonumber
\end{eqnarray}
\begin{eqnarray}
&&
M_{W_H^{\pm}}^2 = m_W^2\left( \frac{f^2}{s^2c^2v^2}-1\right) \,, \nonumber \\
&&
M_{Z_{H}}^{2}=m_{Z}^{2}C_{W}^{2}\left(\frac{f^{2}}{s^{2}c^{2}v^{2}}-1
-\frac{\chi_{H}S_{W}^{2}}{s^{\prime2}c^{\prime2}C_{W}^{2}}\right) \,, \\
&&
M_{A_{H}}^{2}=m_{Z}^{2}S_{W}^{2}\left(\frac{f^{2}}{5s^{\prime2}c^{\prime2}v^{2}}-1
+\frac{\chi_{H}C_{W}^{2}}{4s^{2}c^{2}S_{W}^{2}}\right) \,,~~~~~~~~~~~~~ \nonumber
\end{eqnarray}
with
\begin{eqnarray}
\chi=\frac{4fv^{\prime}}{v^{2}}, \hspace{1cm}
\chi_{H}=\frac{5S_{W}C_{W}}{2}\frac{scs^{\prime}c^{\prime}(c^{2}s^{\prime2}
+s^{2}c^{\prime2})}{5C_{W}^{2}s^{\prime2}c^{\prime2}-S_{W}^{2}s^{2}c^{2}},
\end{eqnarray}
where $m_Z = gv/(2C_W)$, $C_W = \cos\theta_W=\frac{m_w}{m_z}$, $\theta_{W}$ is the Weinberg angle, and  $v^{\prime}$ and $v$ are the VEVs of the scalar $SU(2)_{L}$ triplet and doublet, respectively.

\vskip 5mm
\section{CALCULATION SETUP}
\par
In this section, we present the configuration of the calculation. First, we give a quick overlook of the LO and the NLO calculations, then recapitulate formulism about the transverse momentum resummation and the threshold resummation at the NLL accuracy, for which we refer to Refs.\cite{Fuks:2013vua,Debove:2010kf}. We denote the inclusive hard-scattering $HZ$ production process in hadronic collisions as
\begin{equation}
  A(P_A) + B(P_B) \to H(p_3) + Z(p_4) + X \,,
\label{eq:cli}
\end{equation}
where $H$ and $Z$ with four-momenta $p_{3}$ and $p_{4}$ are produced by a collision of the two protons $A$ and $B$ with four-momenta $P_A$ and $P_B$ separately. $X$ denotes the hadronic remnant of the collision.

\par
\subsection{LO and NLO calculations}
\par
\begin{figure*}
\begin{center}
\includegraphics[scale=1.0]{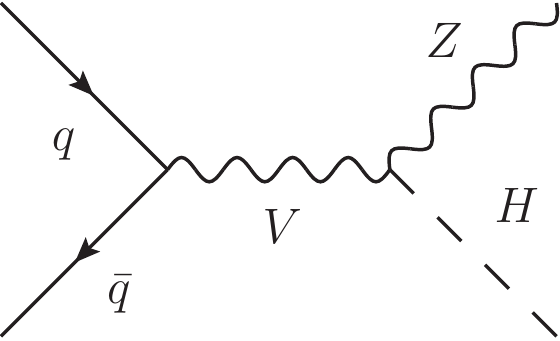}
\caption{ The LO Feynman diagram for the partonic process $q\bar{q}\to HZ$ in the LHM, where $V=Z,Z_H,A_H$, and $q=u,d,c,s,b$. } \label{fig:feyn}
\end{center}
\end{figure*}

At the Born level, the $HZ$ system is produced through
\begin{equation}
q(p_{1}) + \bar q (p_{2})\to H(p_{3}) + Z(p_{4}) \,, (q=u,d,c,s,b),
\label{eq:lo}
\end{equation}
where $p_{1}$ and $p_{2}$ denote the four-momenta of incoming partons. Our calculation shows that the relative difference between the integrated cross sections obtained by adopting $m_b=4.25~{\rm GeV}$ and $m_b=0~{\rm GeV}$ is less than $0.01\%$ for $HZ$ production at the $14~{\rm TeV}$ LHC. That is because of the smallness of the bottom-quark density in the proton compared with other light quarks. Thus we ignore all the quark masses of the $u$, $d$, $c$, $s$, and $b$ quark in our calculations. It can be estimated that the LO cross section of the subprocess (\ref{eq:lo}) is of order $\alpha_{ew}^2$. From the Feynman diagram of the LO subprocesses in Fig.\ref{fig:feyn}, we can see that the cross section for $q\bar q \to HZ$ in the LHM contains potential resonant contributions due to the diagrams with exchange of heavy gauge bosons, $Z_H$ or $A_H$. To dispose of
the singularities due to these resonances, the decay widths of $Z_H$ and $A_H$ are introduced. We adopt the unitary gauge, and the other calculation details can be found in Ref.\cite{Zhang:2012vq}.

\par
Our total NLO QCD correction includes the pure NLO QCD correction and the additional contribution from the one-loop-induced $gg$-fusion channel, where the pure NLO QCD correction to the \ppqqHZ process consists of the following contributions: the virtual corrections and the corresponding renormalization counterterms, the real gluon and real light-quark emission corrections, and the contributions of parton distribution function (PDF) counterterms that absorb part of the collinear singularities of the real gluon and real quark contributions. We use the dimensional regularization method to regularize both the ultraviolet (UV) and the infrared(IR) singularities and adopt the modified minimal substraction ($\overline{MS}$) renormalization scheme. To subtract the IR singularities arising from the real gluon emission contributions, we adopt the two cutoff phase space slicing  method \cite{Harris:2001sx}. The four-momentum of the emitted gluon is denoted as $p_5$. An arbitrary soft cutoff $\delta_{s}$ is introduced to split the phase space of the real gluon emission subprocess into two parts, the soft gluon region ($E_{5} \leq \delta_{s} \sqrt{\hat{s}}/2$) and the hard gluon region ($E_{5} > \delta_{s} \sqrt{\hat{s}}/2$). In addition, another cutoff $\delta_{c}$ is introduced to separate the hard gluon region into a hard collinear ($HC$) region ($\hat{s}_{15}$ or $\hat{s}_{25} \leq \delta_{c}\hat{s}$) and a hard noncollinear ($\overline{HC}$) region ($\hat{s}_{15}$ and $\hat{s}_{25} > \delta_{c}\hat{s}$) where $\hat{s}_{ij} = (p_i+p_j)^2$. The real light-quark emission subprocesses are treated similarly.

\par
We also adopt the dipole subtraction\cite{CS} methods to deal with the IR singularities, and find perfect agreement between the two results. We also checked the NLO QCD corrected total cross section for the $HZ$ production in the SM by comparing the results obtained using our programs and MadGraph package \cite{Alwall:2011uj} separately, and the two calculations agree with each other very well.

\par
\begin{figure*}
\begin{center}
\includegraphics[scale=1]{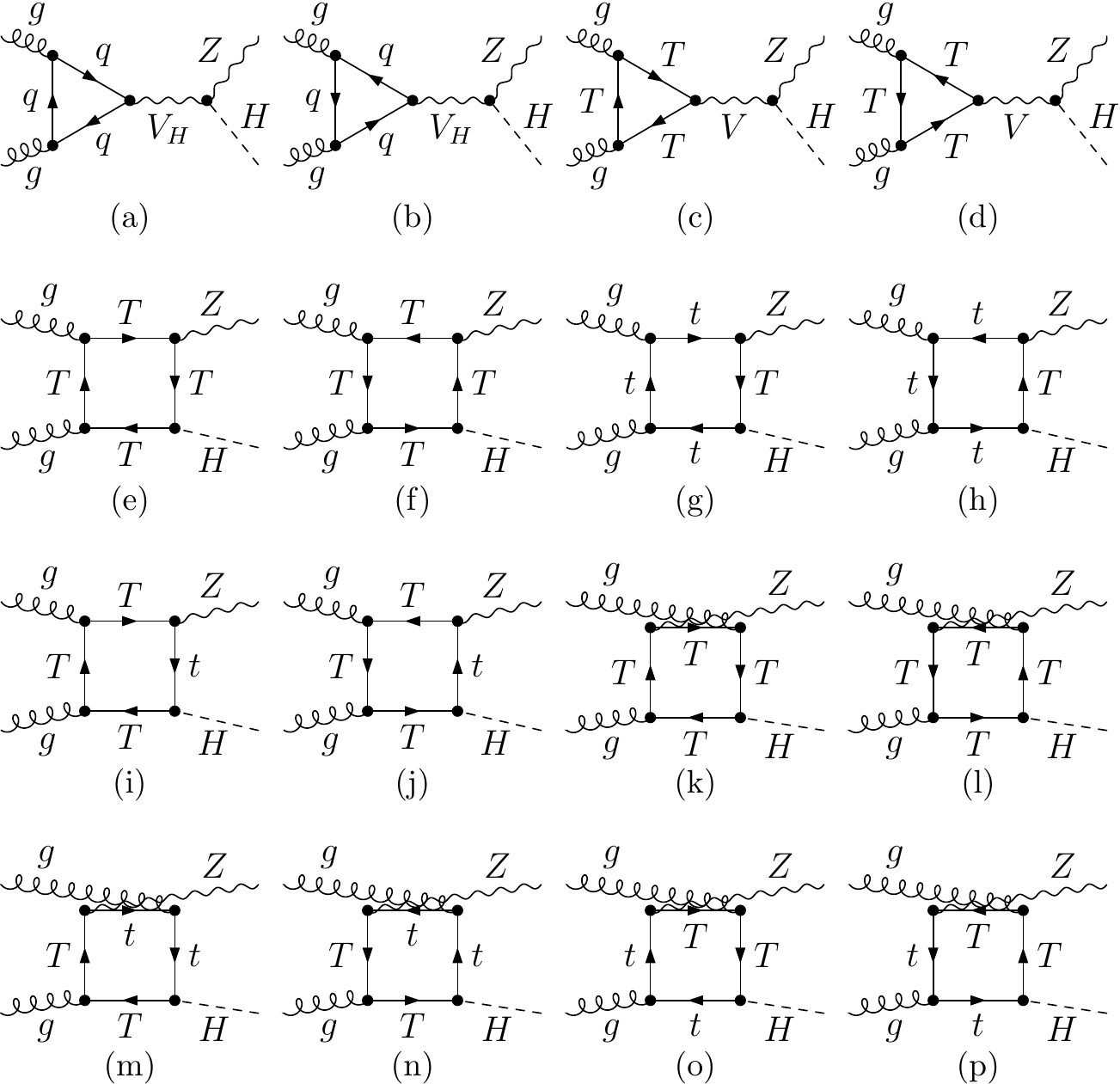}
\caption{ The additional Feynman diagrams in the LHM of the gluon-gluon fusion $gg \to HZ$ subprocess, where $V_H=Z_H,A_H$, $V=Z,Z_H,A_H$, $q=u,d,c,s,t,b$, and $T$ represents the extra top quark partner introduced in the LHM. The SM-like Feynman diagrams for the $gg \to HZ$ subprocess and the diagrams with exchanging of the external gluons are not shown. } \label{fig:gg-HZgraph}
\end{center}
\end{figure*}

Although the cross section at the lowest order for the loop-induced gluon-gluon fusion subprocess $gg \to HZ$ is of $\alpha_{ew}^2\alpha_{s}^2$ order,  of which $\alpha_s$ is an order higher than the QCD NLO contribution from the $q\bar q \to HZ$ subprocess, the former contribution is non-negligible due to the high luminosity of the gluon at the LHC. From Ref.\cite{Brein:2003wg} we know also that with $M_H=125~\text{GeV}$, the NNLO QCD correction to the Drell-Yan channel $q \bar{q} \to HZ$ at the $14~\text{TeV}$ LHC increases the $K$ factor by a mere $1\%$, while the $K$-factor enhancement from the $gg \to HZ$ channel is about $10\%$. Consequently, we include the lowest contribution from the $gg \to HZ$ subprocess within the SM and the LHM in the QCD corrected total cross sections and kinematic distributions for the \ppHZ process, but ignore the other QCD NNLO corrections. The additional Feynman diagrams in the LHM are plotted in Fig.\ref{fig:gg-HZgraph} except for the analogical diagrams in the SM. In the additional one-loop diagrams in the framework of the LHM there is included the internal heavy gauge boson ($A_H$, $Z_H$) and the top quark partner $T$. The total one-loop amplitude ${\cal M}^{1-loop}_{gg}$ is IR and UV finite. The detailed calculation of the gluon-gluon-induced contribution is similar to the analogical evaluation in Ref.\cite{Liang-Wen:2014fla}. We checked the total cross section for the $pp \to gg \to HZ+X$ process in the SM at the $14~{\rm TeV}$ LHC by using our programs and MadGraph separately, and find the results agree with each other.

\par
\subsection{Resummation formalisms }
\par
We denote $M$ and $p_T$ as the invariant mass and transverse momentum of the $HZ$ system, respectively.
By means of the QCD factorization theorem, the inclusive double-differential cross section for the \ppHZ process can be written as \cite{Fuks:2013vua}
\begin{eqnarray}
M^2\frac{\mathrm{d}^2\sigma_{AB}}{\mathrm{d}M^2\mathrm{d}p_T^2}(\tau)
=\sum_{ab}\int_0^1\mathrm{d}x_a\mathrm{d}x_b
\mathrm{d}z \;f_{a/A}(x_a,\mu_F^2)f_{b/B}(x_b,\mu_F^2)\delta\Big(z-\frac{\tau}{x_ax_b}\Big)\Big[z\mathrm{d}\hat{\sigma}_{ab}(z,M^2,p_T^2,\mu_F^2)\Big],~~
\label{eq:diff}
\end{eqnarray}
where $p_{a,b}=x_{a,b}P_{A,B}$, $f_{a/P}(x,\mu_F^2)$ ($a=u,d,c,s,b$) is the PDF of proton, which describes the probability to find a parton $a$ with momentum fraction $x_a$ in proton $P$ at the factorization scale $\mu_F$. $\hat{\sigma}_{ab}$ is the partonic cross section. $\tau=M^2/S$ ($S$ is the hadronic center-of-mass energy squared) and $z=M^2/\hat{s}$ ($\hat{s}$ is the partonic center-of-mass energy squared). We define the Mellin moments of the quantities $F = \sigma_{AB}$, $\hat{\sigma}_{ab}$, $f_{a/A}$ and $f_{b/B}$ through the Mellin transform
\begin{equation}
F(N)=\int_0^1y^{N-1}F(y)\mathrm{d}y \,,
\label{eq:mellin}
\end{equation}
with $y=\tau,z,x_{a}$, and $x_{b}$, respectively. We can rewrite the differential cross section Eq.(\ref{eq:diff}) in Mellin $N$ space as
\begin{equation}
M^2\frac{\mathrm{d}^2\sigma_{AB}}{\mathrm{d}M^2\mathrm{d}p_T^2}(N-1)=
\sum_{ab}f_{a/A}(N,\mu_F^2)f_{b/B}(N,\mu_F^2)\hat{\sigma}_{ab}(N,M^2,p_T^2,\mu_F^2,\mu_R^2) \,.
\label{eq:diff-mellin}
\end{equation}
Under the form of Eq.(\ref{eq:diff-mellin}), we can carry out the resummations of the large logarithmic terms arising in the small transverse momentum and/or the production threshold regions up to all orders in $\alpha_s$ effectively.

\par
\subsubsection{NLL transverse momentum resummation}
\par
In order to take resummation  for the large logarithmic contributions arising at small $p_T$ region, while not violating the transverse momentum conservation, the transverse momentum resummation procedure has to be achieved in the impact-parameter space\cite{Catani:2000vq}. Therefore, a Bessel transform should be applied. The partonic cross section at NLL accuracy in Eq.(\ref{eq:diff-mellin}) can then be expressed by performing the inverse Bessel transform with respect to the impact-parameter $b$ as
\begin{equation}
\hat{\sigma}_{ab}^{NLL}(N,M^2,p_T^2,\mu_F^2,\mu_R^2)=\int_{0}^{\infty}\mathrm{d}b
\frac{b}{2}J_0(bp_T)\times\hat{\sigma}_{ab}^{NLL}(N,M^2,b^2,\mu_F^2,\mu_R^2) \,,
\label{eq:impact}
\end{equation}
where $J_0$ is the zeroth-order Bessel function. The impact-parameter $b$ and $p_{T}$ are conjugated variables. Up to the NLL, the resummed partonic cross section in the $(N,b)$ space can be expressed as \cite{Fuks:2013vua}
\begin{eqnarray}
\hat{\sigma}_{ab}^{NLL}(N,M^2,b^2,\mu_F^2,\mu_R^2) &=& \sum_{a',a'',b',b''}
U_{a'a}^{(1)}(N,1/\bar{b}^2,\mu_F^2)U_{b'b}^{(1)}(N,1/\bar{b}^2,\mu_F^2)
{\cal C}_{a''a'}(N,1/\bar{b}^2) \nonumber \\
&\times& {\cal C}_{b''b'}(N,1/\bar{b}^2){\cal H}_{a''b''}(M^2,\mu_R^2)
\mathrm{exp}[{\cal G}_{a''b''}(\bar{b}^2,M^2,\mu_R^2)] \,,
\label{eq:ptexp}
\end{eqnarray}
where $U_{a'a}^{(1)}$ are evolution operator matrices that evolve the PDFs from the scale
$\mu_F$ to the scale $1/\bar{b}$ with $\bar{b}=be^{\gamma_E}/2$ ($\gamma_E$ is the Euler numb
er).\footnote{The introduction of $e^{\gamma_E}/2$ is to simplify the algebraic expression of $\cal G$ and the
choice is purely conventional\cite{Bozzi:2005wk}.} The hard function ${\cal H}_{ab}(M^2,\mu_R^2)$ is independent of the impact parameter and can be expanded in powers of $\alpha_s$. There are freedoms to separate different contributions into various ${\cal H}_{ab}$, ${\cal G}_{ab}$, and ${\cal C}_{ab}$ functions, which reflects the choice of the resummation scheme \cite{Bozzi:2005wk}. As recommended by Ref.\cite{Fuks:2013vua}, we choose the ``physical'' resummation scheme where the function ${\cal H}_{ab}$ is free from any logarithmic terms and ${\cal G}_{ab}$ and ${\cal C}_{ab}$ are free from any hard contributions, which means they are both universal functions. At the NLL accuracy, the hard function is expressed as
\begin{equation}
{\cal H}_{ab}(M^2,\mu^2_R)=\hat{\sigma}_{ab}^{(0)}(M^2)\left[1+\frac{\alpha_s}{2\pi}{\cal
A}_0\right] \,,
\label{eq:HardF}
\end{equation}
where $\hat{\sigma}_{ab}^{(0)}$ is the Born cross section and ${\cal A}_0$ is the IR-finite part of the renormalized virtual contribution. The expression of ${\cal A}_0$ can be read out from
\begin{eqnarray}
 \hat{\sigma}_{ab}^{V}(M^2,\mu_R^2) &=&
 \frac{\alpha_s(\mu_R^2)}{2\pi}\left(\frac{4\pi\mu_R^2}{M^2}
 \right)^{\epsilon}\frac{\Gamma(1-\epsilon)}{\Gamma(1-2\epsilon)}\hat{\sigma}_{ab}^{(0)}(M^2)
 \nonumber \\
 &\times& \left[ \frac{{\cal A}_{-2}}{\epsilon^2}+ \frac{{\cal
 A}_{-1}}{\epsilon} + {\cal A}_0 \right] \,.
\label{eq:A0}
\end{eqnarray}
At the NLL accuracy, the universal functions ${\cal C}_{ab}$ appearing in Eq.(\ref{eq:ptexp}) are expressed as
\begin{equation}
{\cal C}_{ab}(N,\mu_R^2)=\delta_{ab}+\frac{\alpha_s(\mu_R^2)}{2\pi}\left[\frac{\pi^2}{6}C_a
\delta_{ab}-\gamma_{ab}^{(1),\epsilon}(N)\right] \,,
\label{eq:Ccoe}
\end{equation}
where the QCD color factors are $C_q=C_F$ and $C_g=C_A$ , and $\gamma_{ab}^{(1),\epsilon}(N)$ represent the ${\cal O}(\alpha_s,\epsilon)$ parts of the Altarelli-Parisi splitting kernels in Mellin space. As mentioned above, the Sudakov form factor ${\cal G}_{ab}$ in Eq.(\ref{eq:ptexp}) is chosen to be free from any hard contribution. At the NLL accuracy, it can be expanded as \cite{Bozzi:2006fw,Debove:2009ia}
\begin{eqnarray}
  {\cal G}_{ab}(\bar{b}^2,M^2,\mu_R^2) &=&\ g_{ab}^{(1)} (\lambda)\mathrm{ln}\left(M^2{\bar{b}}^2\right)+
    g_{ab}^{(2)} \Big(\lambda, \frac{M^2}{\mu_R^2} \Big) \ ,
\label{eq:G}
\end{eqnarray}
where $\lambda=\beta_0\mathrm{ln}\left(M^2{\bar{b}}^2\right)\alpha_s/(2\pi)$. The explicit expressions for ${\cal G}_{ab}$ can be found in Ref.\cite{Fuks:2013vua}. The first term in this expansion collects the leading logarithmic contributions,
\begin{eqnarray}
\label{eq:ptg1}
 g_{ab}^{(1)}(\lambda) = \frac{1}{2\lambda\beta_0}
   (A_a^{(1)}+A_b^{(1)}) \big[\lambda+\ln(1-\lambda)\big] \ ,
\end{eqnarray}
and the second term is the next-to-leading pieces written as
\begin{eqnarray}
\label{eq:ptg2}
 g_{ab}^{(2)}(\lambda,M^2/\mu^2_R) &=&
  \frac{1}{2\beta_0} \Big[B_a^{(1)}+B_b^{(1)}\Big]  \ln(1-\lambda)
  + \frac{1}{2\beta_0} \Big[A_a^{(1)}+A_b^{(1)}\Big]\Big[\frac{\lambda}{1-\lambda} +
      \ln(1-\lambda)\Big] \ln\frac{M^2}{\mu_R^2}  \nb \\
 &+& \frac{\beta_1}{2 \beta_0^3} \Big[A_a^{(1)}\!+\!A_b^{(1)}\Big] \Big[
  \frac{\lambda+\ln(1-\lambda)}{1-\lambda}  +\frac{1}{2}\ln^2(1-\lambda)\Big] \nb \\
 &-& \frac{1}{2 \beta_0^2} \Big[A_a^{(2)}+A_b^{(2)}\Big] \Big[\frac{\lambda}{1-\lambda}
  +\ln(1-\lambda)\Big] \ ,
\end{eqnarray}
where the relevant coefficients of the resummation functions $A_a$ and $B_a$ have been expressed as
\begin{eqnarray}
 && A_a^{(1)} = 2C_a \ , \qquad
  A_a^{(2)} = 2C_a\bigg[\bigg(\frac{67}{18}- \frac{\pi^2}{6}\bigg)C_A -\frac{5}{9}N_f\bigg] \ , \nb \\
 &&   B_q^{(1)} = -3 C_F, \qquad~~~~\qquad B_g^{(1)} = -2\beta_0 \ .
\label{eq:A1A2B1}
\end{eqnarray}
Here and in further expressions the associated one-loop coefficient $\beta_0$ and the two-loop coefficient $\beta_1$ are defined by
\begin{equation}
\label{beta1}
  \beta_0 = \frac{11}{6} C_A - \frac23 N_f \tau_R,~~~~~ \beta_1 =  \frac16 \Big[17 C_A^2 - 5C_A N_f - 3 C_F N_f\Big] ,
\end{equation}
where $N_f$ active quark flavors $C_A = 3$, $C_F=4/3$, and $\tau_R=1/2$.

\par
In the interest of obtaining the resummed result in the physical $p_T$ space, we adopt the minimal prescription of Ref.\cite{inversem} for the inverse Mellin transform and the prescription presented in Ref.\cite{inverseb} for the inverse Bessel transform.

\par
In order to avoid double counting of the logarithmic terms in QCD NLO and QCD NLL calculation and to obtain faithful results in all kinematical regions, the summation of the QCD NLO corrected distribution, $d{\sigma}_{AB}^{NLO}/dp_T$, and QCD NLL resummed distribution, $d{\sigma}_{AB}^{NLL}/dp_T$, have to be consistently subtracted by the overlap part $d{\sigma}_{AB}^{\mathrm{overlap}}/dp_T$, i.e.,
\begin{eqnarray}
\frac{{d\sigma}_{AB}^{NLO+NLL}}{dp_T} = \frac{{d\sigma}_{AB}^{NLO}}{dp_T} +
\frac{{d\sigma}_{AB}^{NLL}}{dp_T} - \frac{{d\sigma}_{AB}^{\mathrm{overlap}}}{dp_T} \,,
\label{eq:match}
\end{eqnarray}
which we call the QCD NLO+NLL corrected distribution. In the above equation, the NLL resummed contribution $d{\sigma}_{AB}^{NLL}/dp_T$ is obtained after inserting Eq.(\ref{eq:impact}) into Eq.(\ref{eq:diff-mellin}) and performing relevant integration and transforms. The $d{\sigma}_{AB}^{\mathrm{overlap}}/dp_T$ is obtained by expanding the NLL resummed contribution to fixed order of  $\alpha_s$.

\par
\subsubsection{NLL threshold resummation}
\par
In the threshold region, the partonic cross section in Eq.(\ref{eq:diff-mellin}) can be refactorized into an exponential form at NLL accuracy as
\begin{eqnarray}
\hat{\sigma}_{ab}^{NLL}(N,M^2,\mu_F^2,\mu_R^2) &=&
\sum_{a',b'}U_{a'a}^{(1)}(N,M^2/\bar{N}^2,\mu_F^2)U_{b'b}^{(1)}(N,M^2/\bar{N}^2,\mu_F^2)
\nonumber \\
&\times& \tilde{\cal H}_{a'b'}(M^2,\mu_R^2)\mathrm{exp}[\tilde{\cal G}_{a'b'}(\bar{N}^2,M^2,\mu_R^2)]\,,
\label{eq:thr}
\end{eqnarray}
where the transverse momentum has been integrated over and $\bar{N}=Ne^{\gamma_E}$. The one-loop approximation of the QCD evolution operator $U^{(1)}_{ab}$ drives the behavior of the parton-into-parton density functions with the energy and encompasses collinear radiation \cite{Furmanski:1981cw}. The hard function $\tilde{\cal H}_{ab}$ and the Sudakov form factor $\tilde{\cal G}_{ab}$ can be computed perturbatively. Recently we learned that Eq.(\ref{eq:thr}) at the NLL accuracy can be improved by applying the collinear improvement procedure \cite{Debove:2010kf}, which includes and resums the subleading terms coming from the universal collinear radiation of the initial state partons at the NLL \cite{Kramer:1996iq,Catani:2001ic,Kulesza:2002rh,Almeida:2009jt}. In Eq.(\ref{eq:thr}) we have already applied the collinear improvement procedure \cite{Debove:2010kf}. The hard function $\tilde{\cal H}_{ab}$ and the Sudakov form factor $\tilde{\cal G}_{ab}$ at the NLL accuracy are expressed as
\begin{eqnarray}\label{eq:thr2-1}
\tilde{\cal H}_{ab}(M^2,\mu_R^2)&=&\tilde{\cal H}_{ab}^{(0)}(M^2,\mu_R^2)+\frac{\alpha_s}{2\pi} \tilde{\cal H}_{ab}^{(1)}(M^2,\mu_R^2)\,, \nonumber \\
\tilde{\cal G}_{ab}(N,M^2,\mu_R^2)&=&\tilde{g}_{ab}^{(1)}(\lambda)\mathrm{ln}\bar{N}+
\tilde{g}_{ab}^{(2)}\Big(\lambda,\frac{M^2}{\mu_R^2}\Big) \,,
\label{eq:thr2-2}
\end{eqnarray}
where $\lambda=\beta_0\mathrm{ln}\bar{N}\alpha_s/(2\pi)$. The LO and NLO parts of the ${\cal H}_{ab}$ function read
\begin{eqnarray}
\tilde{\cal H}_{ab}^{(0)}(M^2,\mu_R^2)&=&\hat{\sigma}_{ab}^{(0)}(M^2), \nonumber \\
\tilde{\cal H}_{ab}^{(1)}(M^2,\mu_R^2)&=&\hat{\sigma}_{ab}^{(0)}(M^2)\left[{\cal A}_0+
\frac{\pi^2}{6}(A_a^{(1)}+A_b^{(1)})\right] \,.
\label{eq:thre-H}
\end{eqnarray}
The arguments of the leading and next-to-leading logarithmic contributions to the Sudakov form factor $\tilde{\cal G}_{ab}$ depend, in addition to the reduced Mellin variable, on the one-loop coefficient of the QCD beta function $\beta_0$ which is given as in Eq.(\ref{beta1}).

\par
The coefficients $\tilde{g}^{(1)}_{ab}$ and $\tilde{g}^{(2)}_{ab}$ of the function $\tilde{\cal G}_{ab}$ in Eq.(\ref{eq:thr2-2}) include the resummations of the leading and next-to-leading logarithmic contributions from soft and collinear radiations. In the $\overline{\rm MS}$ renormalization scheme, they are explicitly given
by \cite{Debove:2010kf}
\begin{eqnarray}
  \tilde{g}_{ab}^{(1)}(\lambda)  &=& \frac{1}{2\lambda\beta_0}
    \Big[A_{a}^{(1)}+A_{b}^{(1)}\Big] \Big[2\lambda+\ln(1-2\lambda)\Big] \ , \\
  \tilde{g}_{ab}^{(2)}\Big(\lambda,\frac{M^2}{\mu_R^2}\Big) &=&
     - \frac{1}{2 \beta_0^2} \Big[A_{a}^{(2)} + A_{b}^{(2)}\Big]
      \Big[2\lambda+\ln(1-2\lambda)\Big] \nb \\
  &+& \frac{1}{\beta_0} \Big[B_{a}^{(1)} \!+\! B_{b}^{(1)}\Big] \ln(1-2\lambda)  \nb \\
  &+&  \frac{1}{2\beta_0}\Big[A_{a}^{(1)} + A_{b}^{(1)}\Big]
    \Big[2\lambda+\ln(1-2\lambda)\Big] \ln\frac{M^2}{\mu_R^2} \nb  \\
  &+&  \frac{\beta_1}{2 \beta_0^3}\Big[A_{a}^{(1)} \!+\! A_{b}^{(1)}\Big]
    \Big[2\lambda+ \ln(1\!-\!2\lambda) \!+\! \frac{1}{2}\ln^2(1 - 2\lambda)\Big] \ .
\end{eqnarray}
There the relevant coefficients of the resummation functions $A_a$ and $B_a$ are already expressed in Eq.(\ref{eq:A1A2B1}).

\par
To obtain results in the invariant mass space, the inverse Mellin transform needs to be applied to Eq.(\ref{eq:match}). We still choose the minimal prescription in Ref.\cite{inversem} for the inverse Mellin transform. In analogy to the QCD NLO+NLL corrected transverse momentum distribution, the QCD NLO+NLL corrected invariant mass distribution is obtained as
\begin{eqnarray}
\frac{{d\sigma}_{AB}^{NLO+NLL}}{dM} = \frac{{d\sigma}_{AB}^{NLO}}{dM} +
\frac{{d\sigma}_{AB}^{NLL}}{dM} - \frac{{d\sigma}_{AB}^{\mathrm{overlap}}}{dM}.
\label{eq:mmatch}
\end{eqnarray}
In the above equation, the NLL resummed contribution $d{\sigma}_{AB}^{NLL}/dM$ is obtained by performing the integration over $p_T$ for Eq.(\ref{eq:diff-mellin}), inserting Eq.(\ref{eq:thr}) into Eq.(\ref{eq:diff-mellin}), and performing inverse Mellin transform. $d{\sigma}_{AB}^{\mathrm{overlap}}/dM$ is obtained by expanding the NLL resummed contribution to the order of $\alpha_s$. From Eq.(\ref{eq:mmatch}) we can also obtain the QCD NLO+NLL corrected total cross section after performing the integration over $M$.

\vskip 5mm
\section{NUMERICAL ANALYSIS}
\par
\subsection{Input parameters}
\par
The input parameters in the numerical calculations are as follows. The $q$ quarks ($q=u,d,c,s,b$) are taken as massless. We used the $G_{\mu}$ scheme for the fine-structure constant, i.e., the electromagnetic coupling constant $\alpha$ is derived from the Fermi coupling constant $\alpha_{G_{\mu}}= \sqrt{2}G_{\mu} M^2_W (1- M^2_W/M^2_Z)/\pi$. The SM parameters are taken as \cite{Agashe:2014kda}
\begin{align}
&   M_W = 80.385 \,\text{GeV} ,  \quad \quad   M_Z = 91.1876 \, \text{GeV} \, ,  \nonumber  \\
&   M_t = 173.21 \, \text{GeV},  \quad \quad  G_{\mu} = 1.1663787\times 10^{-5}\, \text{GeV}^{-2} \, .
\end{align}

\par
The light neutral Higgs mass is taken as $M_H = 125~{\rm GeV}$, and the Weinberg mixing angle in the SM is obtained from $S^2_W=1- M^2_W/M^2_Z$. The vacuum expectation value of the Higgs doublet is chosen as $v=246~{\rm GeV}$. The CT10 and CT10nlo PDFs are adopted in the LO and NLO/NLO+NLL calculations, respectively. The strong coupling constant $\alpha_s$ provided by the CT10 PDFs \cite{Lai:2010vv} is used in the calculation. To make theoretical predictions for integrated cross sections, we take three distinct value sets for the LHM parameters considering the constraints of the electroweak precision data on LHM parameters \cite{Reuter:2012sd,Csakj}. We fix $\chi = 0.5$, $R=1$, and the other input LHM parameters are chosen representatively in three parameter cases, in order to show the effects of these parameters. Namely, (1) case A, $c = 0.5$, $c^{\prime}=0.22$, and $f=4~{\rm TeV}$; (2) case B, $c = 0.3$, $c^{\prime}=0.3$, and $f=4.5~{\rm TeV}$; (3) case C, $c = 0.8$, $c^{\prime}=0.4$, and $f=5~{\rm TeV}$. This analysis  provides us crucial information to test the experimental possibility for the $HZ$ production process in the LHM context. The corresponding heavy gauge bosons and the $T$-even partner of the top quark have masses $M_{Z_H}=3024.1\,{\rm GeV}$, $M_{A_H}= 1462.5\,{\rm GeV}$, and $M_T=5632.8\,{\rm GeV}$ for case A; $M_{Z_H}=5147.5\,{\rm GeV}$, $M_{A_H}= 1232.7\,{\rm GeV}$, and $M_T=6337.0\,{\rm GeV}$ for case B; $M_{Z_H}=3406.1\,{\rm GeV}$, $M_{A_H}= 1068.5\,{\rm GeV}$, and $M_T=7041.1\,{\rm GeV}$ for case C.

\par
\subsection{Total cross section }
\par
We include the one-loop-induced $gg$-fusion channel contribution in the QCD corrected integrated cross sections in both the SM and LHM. The QCD NLO+NLL corrected total cross section is obtained by performing the integration for Eq.(\ref{eq:mmatch}) over $HZ$ invariant mass $M$, and combining with the one-loop-induced $gg$-fusion channel contribution. For simplicity we set the factorization and renormalization scales being equal ($\mu_R=\mu_F=\mu$) and in the total cross section calculation we fix the scale $\mu$ as the central value of $\mu_0=M_Z+M_H$ if there is no other statement. We list the LO, QCD NLO, NLO+NLL corrected total cross sections, and the contributions from the $gg$-fusion partonic process for the $HZ$ production with the three LHM parameter cases at the $14~{\rm TeV}$ LHC in Table \ref{tab:total-1}. We can see from the table that the $gg$-fusion contribution is numerically relevant in the predicted cross section, even more important than the NLL resummation effect in the QCD NLO+NLL calculation. In further calculations and analyses we fix case A values for $c$, $c^{\prime}$, and $f$ parameters.
%%%%%%%%%%%%%%%% Table%%%%%%%%%%%%%%%%%%%%%%%%%%%%%%%%%%%%%%%%%%
\begin{table}

\renewcommand{\arraystretch}{1.5}
\begin{center}
\begin{tabular}{c|cccc}
\toprule
Cross section (pb) & $\sigma_{\text{LO}}$ & $\sigma_{\text{NLO}}$ & $\sigma_{\text{NLO+NLL}}$ & $\sigma_{gg}$ \\
\hline
Case A & $0.807(1)$ & $1.007(1)$ & $1.001(1)$ & $0.072(1)$ \\
\hline
Case B & $0.808(1)$ & $1.010(1)$& $1.004(1)$ & $0.072(1)$\\
\hline
Case C & $0.787(1)$ & $0.989(1)$ & $0.984(1)$ & $0.072(1)$ \\
\bottomrule
\end{tabular}
\end{center}
\caption{The LO, QCD NLO, and NLO+NLL corrected total cross sections predicted in the LHM for the $HZ$ production at the $\sqrt{S}=14\,\text{TeV}$ LHC. The contribution from the $gg$-fusion subprocess is also listed independently. Case A, case B, and case C represent different LHM parameter sets. }
\label{tab:total-1}
\end{table}

\par
\begin{figure*}
\begin{center}
\includegraphics[scale=0.5]{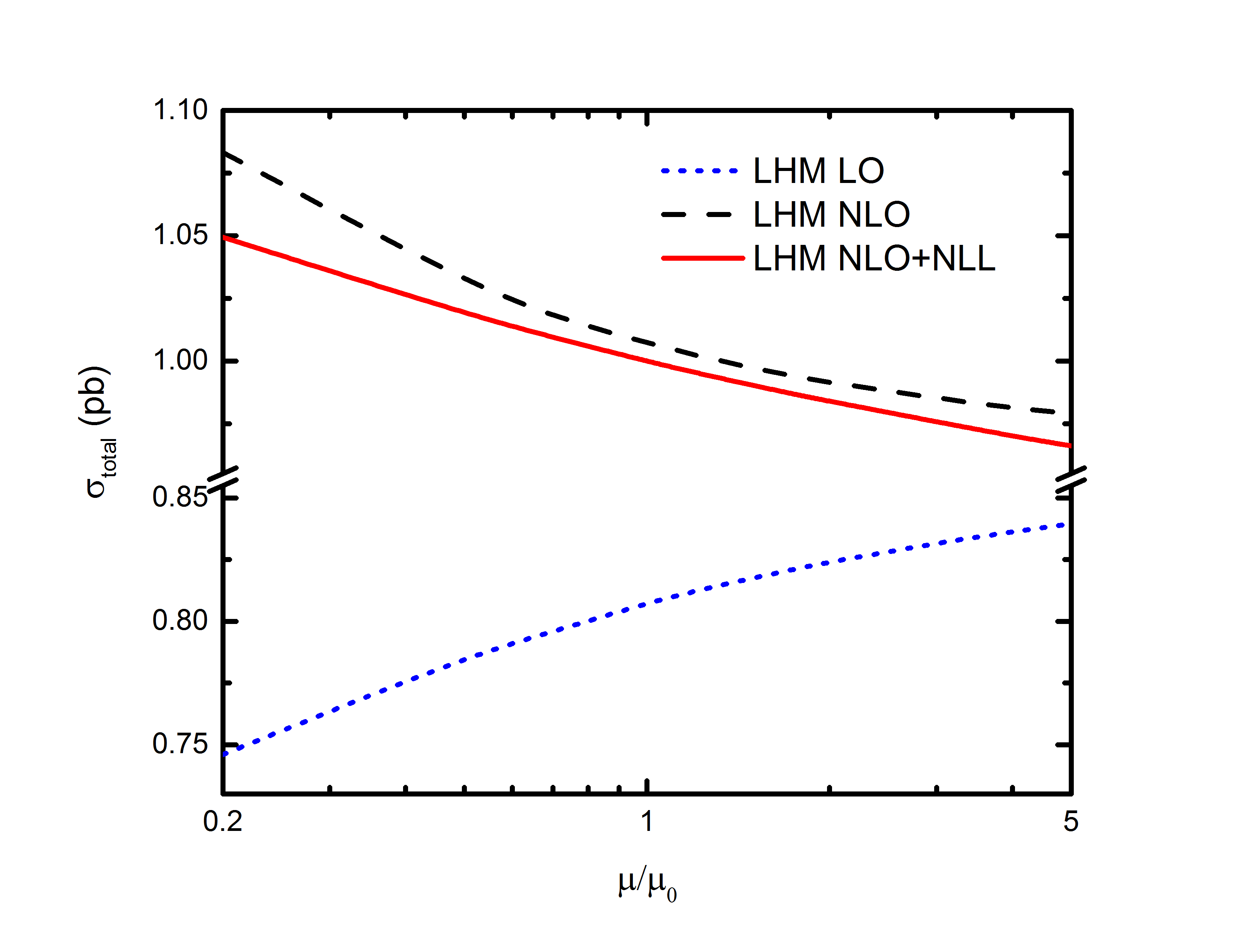}
\caption{ The factorization/renormalization scale dependence of the total cross sections of the $HZ$ production in the LHM at the $14\,\text{TeV}$ LHC. The dotted, dashed, and solid curves are for the LO, NLO, and NLO+NLL, respectively. }
\label{fig:totalth}
\end{center}
\end{figure*}

The LO, QCD NLO, and NLO+NLL corrected integrated cross sections for the $HZ$ production in the LHM at the $14~{\rm TeV}$ LHC as the functions of the factorization/renormalization scale are depicted in Fig.\ref{fig:totalth}, where the scale $\mu$ varies from $0.2$ to $5\mu_0$. The dotted curve is for the LO cross section, the dashed curve is for the NLO QCD corrected cross section, and the solid curve is for the QCD NLO+NLL corrected cross section. Normally for a process involving pure electroweak interaction subprocesses at the LO, one does not expect a significant scale uncertainty improvement at the QCD NLO. But Fig.\ref{fig:totalth} shows clearly that the NLO QCD correction reduces obviously the scale dependence of the total cross section, and the QCD NLO+NLL correction improves the scale uncertainty even better than the pure QCD NLO correction.
%%%%%%%%%%%%%%%%%%%%%%%%%%%%%%%%%%%%%%%%%%%%%%%%%%%%%%%%%%

\par
Except for the theoretical scale uncertainty, there is another uncertainty of the PDF, which is associated with the experimental data adopted to build the PDF fits. The PDF uncertainty is normally not improved by high-order evaluation procedure. The CT10 collaboration uses the Hessian method to estimate the PDF experimental uncertainty by propagating the experimental uncertainties on the fitted data and leads to the production of orthogonal eigenvector PDF sets corresponding to a $90\%$ confidence level \cite{Martin:2009iq}. The PDF errors on the cross section are then obtained by the following formulas,
\begin{eqnarray}
 \Delta\sigma_{\rm PDF+}&=&\sqrt{\sum_{i=1}^{n}\left[\max\left(\sigma_{+i}-\sigma_0,
 \sigma_{-i}-\sigma_0,0\right)\right]^2},\\
 \Delta\sigma_{\rm PDF-}&=&\sqrt{\sum_{i=1}^{n}\left[\max\left(\sigma_0-\sigma_{+i},
 \sigma_0-\sigma_{-i},0\right)\right]^2},
\end{eqnarray}
where the number of eigenvector directions in the CT10 fit is $n=26$, and $\sigma_0$ is the cross section calculated with the best fit PDF set. In our calculations of the PDF uncertainty, we use CT10 PDF sets to figure out the PDF uncertainty as the deviation range of the total cross section.

\par
We list in Table \ref{tab:total} the integrated total cross sections and the corresponding errors in the LHM and the SM at the $14\,\text{TeV}$ LHC. There we list also the cross sections contributed by the one-loop-induced $gg$-fusion partonic process, $\sigma_{LHM}^{gg}$ and $\sigma_{SM}^{gg}$. In the table, the upper and lower errors mean the upper and lower limitations from the scale error and the PDF error, respectively. The scale error limitations are defined by varying $\mu$ from $0.5$ to $2\mu_0$. The central values represent the total cross section with $\mu=\mu_0$. From the data for both the LHM and SM in the table we can see again that the NLO QCD correction reduces the total theoretical error of the total cross section, and the total error is further reduced by including both the QCD NLO and NLL corrections. These numerical results demonstrate that the scale uncertainty of the QCD NLO corrected total cross section is less than the LO one, while the QCD NLO+NLL correction reduces more significantly the scale uncertainty than the QCD NLO correction. Furthermore, we find that the scale uncertainty including the QCD NLO correction is mainly contributed by the lowest order $gg$-fusion subprocess. If discarding the $gg$-fusion correction, the QCD NLO and NLO+NLL scale uncertainties would be decreased further. In this table the LO, QCD NLO, and NLO+NLL relative deviations ($\delta$) of the LHM predicted total cross sections from the corresponding ones in the SM are defined as
\begin{eqnarray}
\label{relative-discripancy}
 \delta_{LO} = \frac{\sigma^{LHM}_{LO}-\sigma^{SM}_{LO}}{\sigma^{SM}_{LO}}, ~~ \delta_{NLO} = \frac{\sigma^{LHM}_{NLO}-\sigma^{SM}_{NLO}}{\sigma^{SM}_{NLO}},~~  \delta_{NLO+NLL} = \frac{\sigma^{LHM}_{NLO+NLL}-\sigma^{SM}_{NLO+NLL}}{\sigma^{SM}_{NLO+NLL}}. \nb  \\
\end{eqnarray}
The relative deviations listed in the table show that the QCD NLO correction reduces $\delta_{LO}$ obviously, while the QCD NLO+NLL correction decreases the NLO relative deviation slightly. We conclude that (1) the theoretical scale+PDF uncertainty of the total cross section can be improved by including both the QCD NLO correction and the NLL threshold resummation; (2) the QCD NLO+NLL correction decreases the relative deviation from the SM total cross sections obviously, but the LHM effect in the $HZ$ production process is still observable after taking the QCD NLO+NLL effects into account in precision study.
%%%%%%%%%%%%%%%% Table%%%%%%%%%%%%%%%%%%%%%%%%%%%%%%%%%%%%%%%%%%
\begin{table}
\renewcommand{\arraystretch}{1.5}
\begin{center}
\begin{tabular}{c|cccc}
\toprule
 Cross section & LO & NLO & NLO+NLL & gg fusion \\
\hline
$\sigma_{LHM}\,(\text{pb})$ & $0.807^{+0.018+0.023}_{-0.022-0.025}$ & $1.007^{+0.023+0.027}_{-0.016-0.027}$ & $1.001^{+0.018+0.027}_{-0.016-0.027}$ & $0.072^{+0.019+0.003}_{-0.014-0.003}$ \\
\hline
$\sigma_{SM}\,(\text{pb})$ & $0.731^{+0.023+0.021}_{-0.029-0.022}$ & $0.927^{+0.023+0.023}_{-0.014-0.026}$ & $0.924^{+0.018+0.023}_{-0.016-0.026}$ & $0.073^{+0.019+0.003}_{-0.014-0.003}$ \\
\hline
$\delta$ & $10.4\%$ & $8.6\%$ & $8.3\%$ & $-1.5\%$ \\
\bottomrule
\end{tabular}
\end{center}
\caption{The LO, QCD NLO, and NLO+NLL corrected total cross sections and the relative deviations of the cross sections predicted in the LHM from that in the SM for the $HZ$ production at the $\sqrt{S}=14\,\text{TeV}$ LHC. The contribution from the $gg$-fusion subprocess is also listed independently. For each result, the central values represent the total cross section obtained by taking $\mu=\mu_0$; the first error is due to scale uncertainty in the scale range of $0.5\mu_0 \le \mu \le 2 \mu_0$, and the second error is due to the PDF uncertainty. }
\label{tab:total}
\end{table}

\par
In Table \ref{tab:cut} we list the total cross sections in the LHM and the corresponding relative deviations after applying a lower cut on $HZ$ invariant mass ($M^{cut}$) to demonstrate the way to promote the possibility for finding LHM evidence. The table shows that in the range of $250\,{\rm GeV} \le M^{cut} \le 400\,{\rm GeV}$ the QCD NLO corrections to the LO cross sections are always positive and the NLO+NLL corrections reduce slightly the corresponding NLO corrected ones. The results of $\delta$ in Table \ref{tab:cut} show that the LO, NLO, and NLO+NLL relative deviations [defined in Eq.(\ref{relative-discripancy})] increase rapidly as the low cut $M^{cut}$ goes up. For example, we can read out that the relative deviation $\delta_{NLO+NLL}$ is about $12.4\%$ for $M^{cut}=250~{\rm GeV}$ and increases to $71.2\%$ for $M^{cut}=400\,{\rm GeV}$. That means the new physics sign of the LHM becomes more obvious if we take a large enough lower cut on the $HZ$ invariant mass.
%%%%%%%%%%%%%%%% Table%%%%%%%%%%%%%%%%%%%%%%%%%%%%%%%%%%%%%%%%%%sacrifice
\begin{table}
\renewcommand{\arraystretch}{1.4}
\begin{center}
\begin{tabular}{c|c|c|c|c}
\toprule
\backslashbox{{cross section}}{{\small $M^{cut}\,(\text{GeV})$}}  &  250 & 300  & 350 & 400  \\  \hline
$\sigma^{LHM}_{LO} (\rm fb)$  & 587.66(5)  & 354.13(1)  & 241.68(1) & 184.49(2)    \\
$\sigma^{SM}_{LO}(\rm fb)$    &  506.96(3) & 266.65(1)  & 150.56(1) &  91.06(1)    \\
$\delta_{LO} $                & $15.9\%$   & $32.8\%$   & $60.5\%$  & $102.6\%$    \\ \hline
$\sigma^{LHM}_{NLO}(\rm fb)$  & 752.8(8)   & 474.9(4)   & 328.3(3)  & 238.6(4)     \\
$\sigma^{SM}_{NLO}(\rm fb)$   & 666.5(7)   & 380.9(1)   & 229.8(1)  & 137.3(2)     \\
$\delta_{NLO}$                & $12.9\%$   & $24.7\%$   & $42.9\%$  & $73.8\%$     \\ \hline
$\sigma^{LHM}_{NLO+NLL}(\rm fb)$ & 746.6(8)& 469.8(4)   & 323.6(3)  & 234.0(4)     \\
$\sigma^{SM}_{NLO+NLL}(\rm fb)$  & 664.3(7)& 379.8(1)   & 229.0(1)  & 136.7(2)     \\
$\delta_{NLO+NLL}$               & $12.4\%$& $23.7\%$   & $41.3\%$  & $71.2\%$     \\
 \bottomrule
\end{tabular}
\caption{The LO, QCD NLO, and NLO+NLL corrected total cross sections and the corresponding relative deviations for the $pp\to HZ+X$ process at the $14~{\rm TeV}$ LHC with different values of the lower cut ($M^{cut}$) on the invariant mass. }
\label{tab:cut}
\end{center}
\end{table}

\par
\subsection{Transverse momentum distribution}
\par
\begin{figure*}
\begin{center}
\includegraphics[scale=0.45]{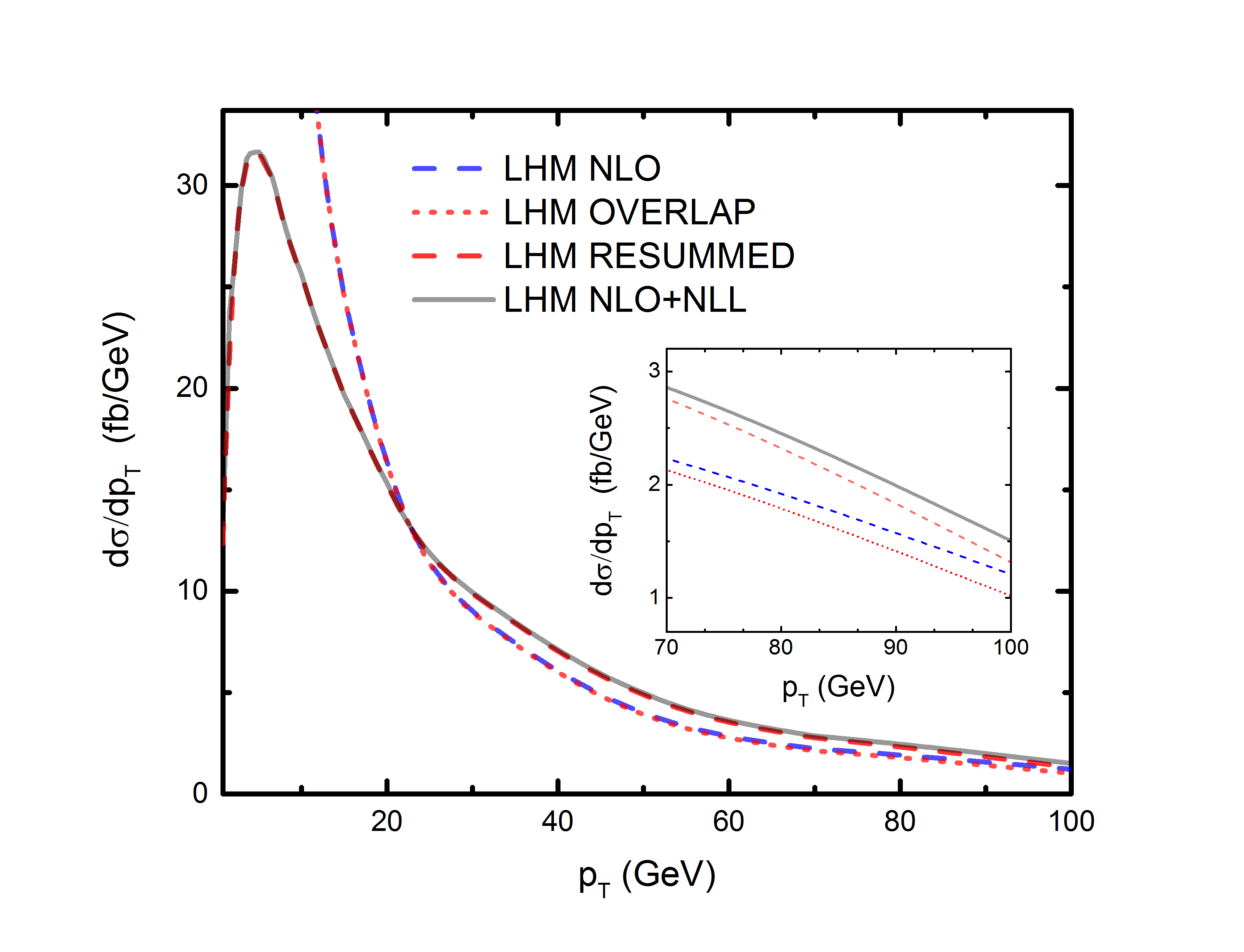}
\caption{ The transverse momentum distributions of the $HZ$ production within the LHM at the $14\,\text{TeV}$ LHC. The QCD NLO corrected distribution is drawn with a blue dashed curve, the overlapped distribution with a red dotted curve, the QCD NLL resummed distribution with a red dashed curve, and the QCD NLO+NLL corrected distribution with a black full curve. }
\label{fig:exp}
\end{center}
\end{figure*}

\label{Sec-4-3}
Now we turn to the transverse momentum distribution of the $HZ$ production within the LHM in the  NLO+NLL QCD. Here we denote the transverse momentum of the $HZ$ system simply as $p_T$. As we know, the ratio of $\sigma_{gg}/\sigma_{NLO}$ is only about $7\%$ as shown in Table \ref{tab:total-1}, and the one-loop-induced $gg$-fusion channel does not provide contribution to the $p_T$ distribution due to the conservation of the transverse momentum of the final $HZ$ system. Therefore, it is justified to consider only the contribution from the dominant $q\bar{q}$ annihilation channel with the NLO+NLL QCD accuracy in the following discussion of $p_T$ distribution. The NLO+NLL QCD corrected $p_T$ distribution is obtained by using Eq.(\ref{eq:match}). In calculating the $p_T$ distributions of the $HZ$ system, we identify the unphysical scale $\mu=\mu_F=\mu_R$ with $\mu_0=M_Z + M_H$ unless there is another statement. In Fig.\ref{fig:exp} we show the QCD NLO corrected, the NLL resummed, the overlapped part, and the QCD NLO+NLL corrected $HZ$ transverse momentum distributions in the LHM at the $14\,{\rm TeV}$ LHC. We can see that the overlapped $p_T$ distribution and the QCD NLO corrected distributions are in good agreement particularly in the low $p_T$ region, but as $p_T$ becomes larger, the discrepancy between the two results becomes more obvious; for example, at $p_T=100\,{\rm GeV}$ the discrepancy reaches $16\%$. We see also that the QCD NLO corrected distribution shows divergence tendency at the low $p_T$ region, while the QCD NLO+NLL corrected distribution exhibits a finite and physical behavior having a peak around $5\,{\rm GeV}$ in low $p_T$ area. From this respect, we can conclude that after taking account of the resummation effects the $p_T$ distribution will be more reliable.
%%%%%%%%%%%%%%%%%%%%%%%%%%%%%%%%%%%%%%%%%%%%%%%%%%%%%%%%%%%%%%%%%%%

\par
\begin{figure*}
\begin{center}
\includegraphics[scale=0.45]{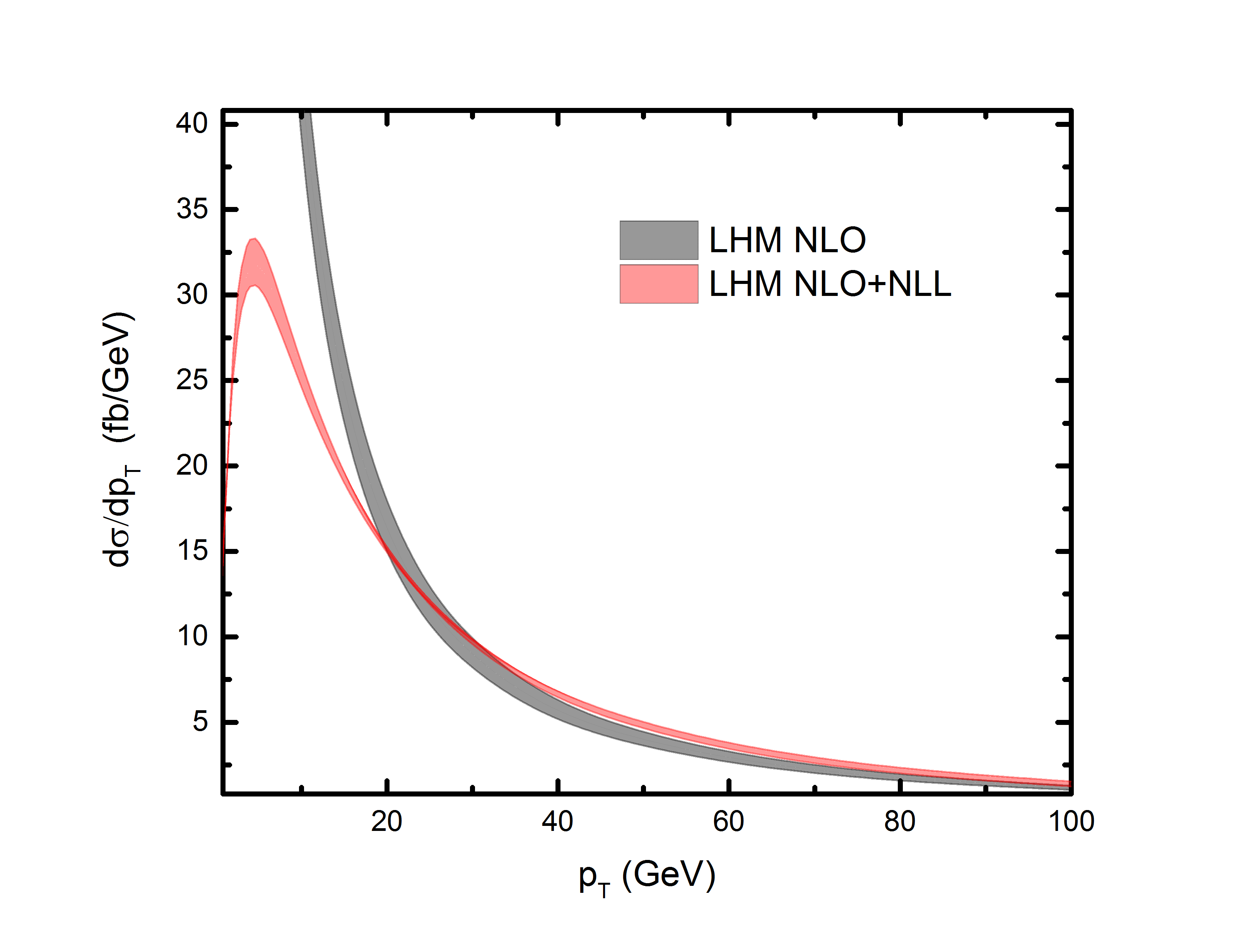}
\caption{ The $HZ$ transverse momentum distributions and the related scale uncertainty of the $HZ$ production in the LHM at the $14~{\rm TeV}$ LHC. The QCD NLO corrected $p_T$ distribution range is shown as the gray band and the QCD NLO+NLL corrected distribution range as the red band with the scale varying in the range of $[\frac{1}{2}\mu_0,~2\mu_0]$. }
\label{fig:mu}
\end{center}
\end{figure*}

To estimate the scale uncertainty of differential cross sections, we define the scale uncertainty in a usual way from the variation of the factorization/renormalization scale, where the scale varies around the central value $\mu_0=M_Z + M_H$ from $\frac{1}{2}$ to $2\mu_0$. In Fig.\ref{fig:mu} we plot the transverse momentum distributions of the $HZ$ production with the corresponding scale uncertainties within the LHM at the $14~{\rm TeV}$ LHC. It shows that the QCD NLO corrected distribution exhibits a much wider band than the QCD NLO+NLL corrected distribution, which means that the QCD NLO+NLL corrected distribution owns a better theoretical scale uncertainty. In Table \ref{tab:ptmu}, we list the results for the relative scale uncertainty for some typical $p_T$ with its definition as
\begin{equation}
\eta(p_T)=\frac{ {\rm max} \left[ \frac{d\sigma}{dp_T}(\mu)\right] - {\rm min}\left[ \frac{d\sigma}{dp_T} (\mu)\right]}{\frac{d\sigma}{dp_T}(\mu_0)},~~~~(\mu \in [\frac{1}{2}\mu_0,~2 \mu_0]).
\label{eq:diff-uncertainty}
\end{equation}
From the table, we can read out $\eta(p_T=15~{\text{GeV}})=18\%$ and $3\%$ and $\eta(p_T=50~{\rm GeV})=21\%$ and $8\%$ for the QCD NLO and NLO+NLL corrected distributions, respectively. The relative scale uncertainty for the QCD NLO corrected $p_T$ distribution is always larger than the QCD NLO+NLL corrected $p_T$ distributions in the listed range. We conclude that the differential cross section of $p_T$ obtained at the QCD NLO+NLL accuracy is much more reliable than those at QCD NLO.
%%%%%%%%%%%%%%%%%%%%%%%%%%%%%%%%%%%%%%%%%%%%%%%%%%%%%%%%%%%%%%%%%%%%
%%%%%%%%%%%%%%%%%%%%%%%%%%%%%%%%%%%%%%%%%%%%%%%%%%%%%%%%%%%%%%%%%%%
\begin{table}
\renewcommand{\arraystretch}{1.2}
\begin{center}
\begin{tabular}{lcccl}
\toprule
$p_T\,(\text{GeV})$ & $\eta_{NLO}\,(\%)$ & $\eta_{NLO+NLL}\,(\%)$ \\ \midrule
5 & 17 & 8 \\
10 & 17 & 6 \\
15 & 18 & 3 \\
20 & 18 & 3 \\
50 & 21 & 8 \\
100 & 21 & 19 \\
\bottomrule
\end{tabular}
\caption{The relative scale uncertainties of the $p_T$ distribution of the $pp\to HZ+X$ process in the LHM
at the $14~{\rm TeV}$ LHC for some typical values of $p_T$. The relative scale uncertainty is defined in Eq.(\ref{eq:diff-uncertainty}).  }
\label{tab:ptmu}
\end{center}
\end{table}

\par
\begin{figure*}
\begin{center}
\includegraphics[scale=0.4]{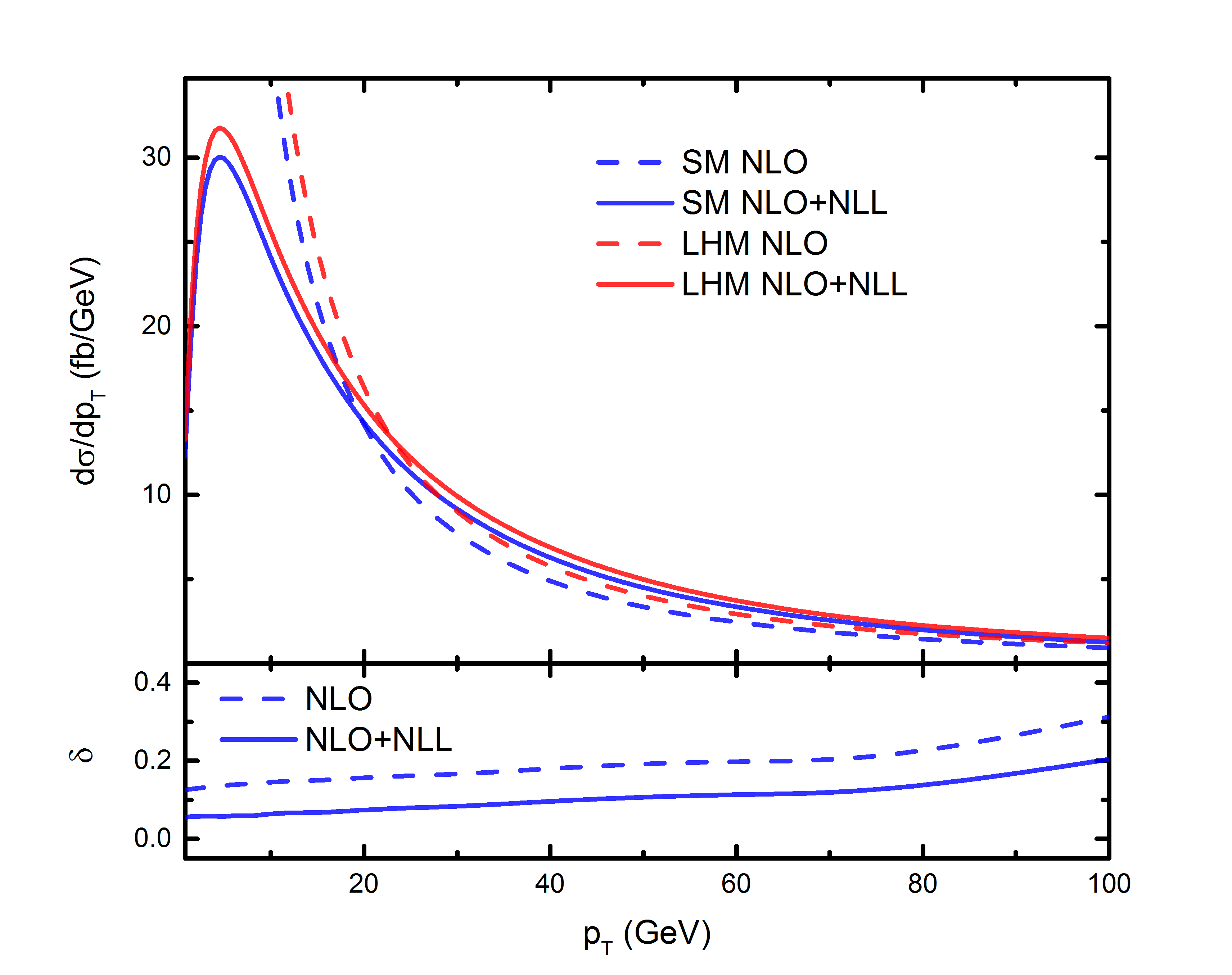}
\caption{\small The $HZ$ transverse momentum distributions for the \ppHZ process in the SM and the LHM (in the upper panel) and the corresponding relative deviations (in the lower panel) at the $14~{\rm TeV}$ LHC. }
\label{fig:ptlh}
\end{center}
\end{figure*}

To describe the relative deviation of the $p_T$ distributions in the LHM from the corresponding SM predictions, we define
\begin{equation}
\delta(p_T)=\frac{\left(\frac{d\sigma}{dp_{T}}\right)_{\text
{\tiny{LHM}}}-\left(\frac{d\sigma}{dp_{T}}\right)_{\text
{\tiny{SM}}}}{\left(\frac{d\sigma}{dp_{T}}\right)_{\text {\tiny{SM}}}}.
\label{eq:delta}
\end{equation}
In Fig.\ref{fig:ptlh}, the upper panel provides the $HZ$ transverse momentum distributions for the $HZ$ production at the $14~{\rm TeV}$ LHC in the LHM and the SM, and the lower panel shows the corresponding relative deviations $\delta(p_T)$. We see from the figure that the QCD NLO corrected $p_T$ distribution in the LHM is larger than that in the SM and both curves share a similar shape. From the upper panel of Fig.\ref{fig:ptlh}, we find that after resummation procedure, the NLO+NLL corrected $p_T$ distributions in both the LHM and the SM are convergent in
the low $p_T$ range as expected. We can see clearly from the lower panel of Fig.\ref{fig:ptlh} that the resummation correction exerts an obvious effect on the $HZ$ transverse momentum distribution. We can read out from the figure that the relative deviation of the QCD NLO corrected distribution varies from $12\%$ to $32\%$ with the increment of $p_T$ in the plotted range, while after resummation $\delta(p_T)$ is evidently reduced to the range of $5\% \sim 20\%$. That implies that the LHM effect on the $HZ$ transverse momentum distribution could be even harder to  measure, but still observable if taking into account the QCD NLO+NLL correction in precision search for the LHM.

\par
\subsection{Invariant mass distribution }
\par
\begin{figure*}
\begin{center}
\includegraphics[scale=0.45]{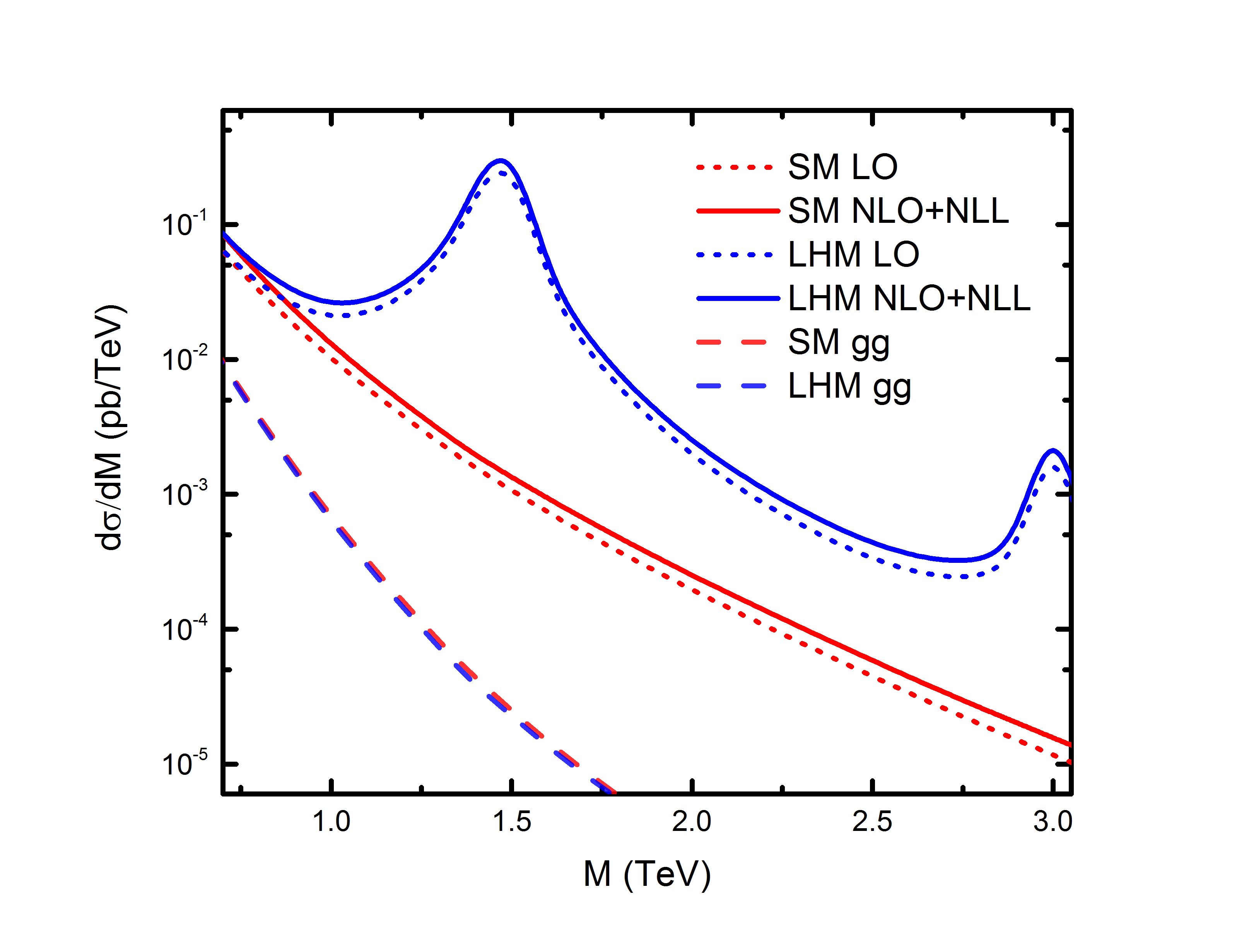}
\caption{\small The LO and NLO+NLL corrected $HZ$ invariant mass distributions for the $HZ$ production in the SM and the LHM at the $14~{\rm TeV}$ LHC. The contribution parts from the $gg$-fusion subprocess among the NLO+NLL corrected distributions are shown independently.}
\label{fig:mlh}
\end{center}
\end{figure*}

In this subsection we discuss the threshold resummation effect on the invariant mass distribution. For the spectra in the invariant mass $M$, we fix the scale to be the invariant mass of the $HZ$ system ($\mu=\mu_F=\mu_R=M_{ZH}$), and denote the $HZ$-system invariant mass as $M$ for simplicity in the following invariant mass distribution analysis. The QCD NLO+NLL corrected invariant mass distribution is obtained via Eq.(\ref{eq:mmatch}) and added together with the contribution from one-loop-induced $gg$-fusion channel. We plot the LO and NLO+NLL corrected $HZ$ invariant mass distributions for the $HZ$ invariant mass in the LHM and SM at the $14~{\rm TeV}$ LHC in Fig.\ref{fig:mlh}, and the corresponding contributions from the one-loop-induced $gg$-fusion subprocess are also plotted independently. We can see that the contributions from the $gg$-fusion channel are much smaller than the corresponding differential cross sections, and their contributing proportions are less than $10\%$ in the plotted range. The figure shows that with the increment of $M$, the LO and NLO+NLL corrected differential cross sections in both the LHM and the SM decrease significantly except in the vicinities of the two resonances for the LHM distributions, i.e., at $M \sim 1500~{\rm GeV}$ and $M \sim 3000~{\rm GeV}$, respectively. Furthermore, the difference between the invariant mass distributions in the LHM and the SM becomes considerably larger, particularly in the two resonant regions, as the invariant mass $M$ grows up.
%%%%%%%%%%%%%%%%%%%%%%%%%%%%%%%%%%%%%%%%%%%%%%%%%%%%%%%%%%%%

\par
\begin{figure*}
\begin{center}
\includegraphics[scale=0.45]{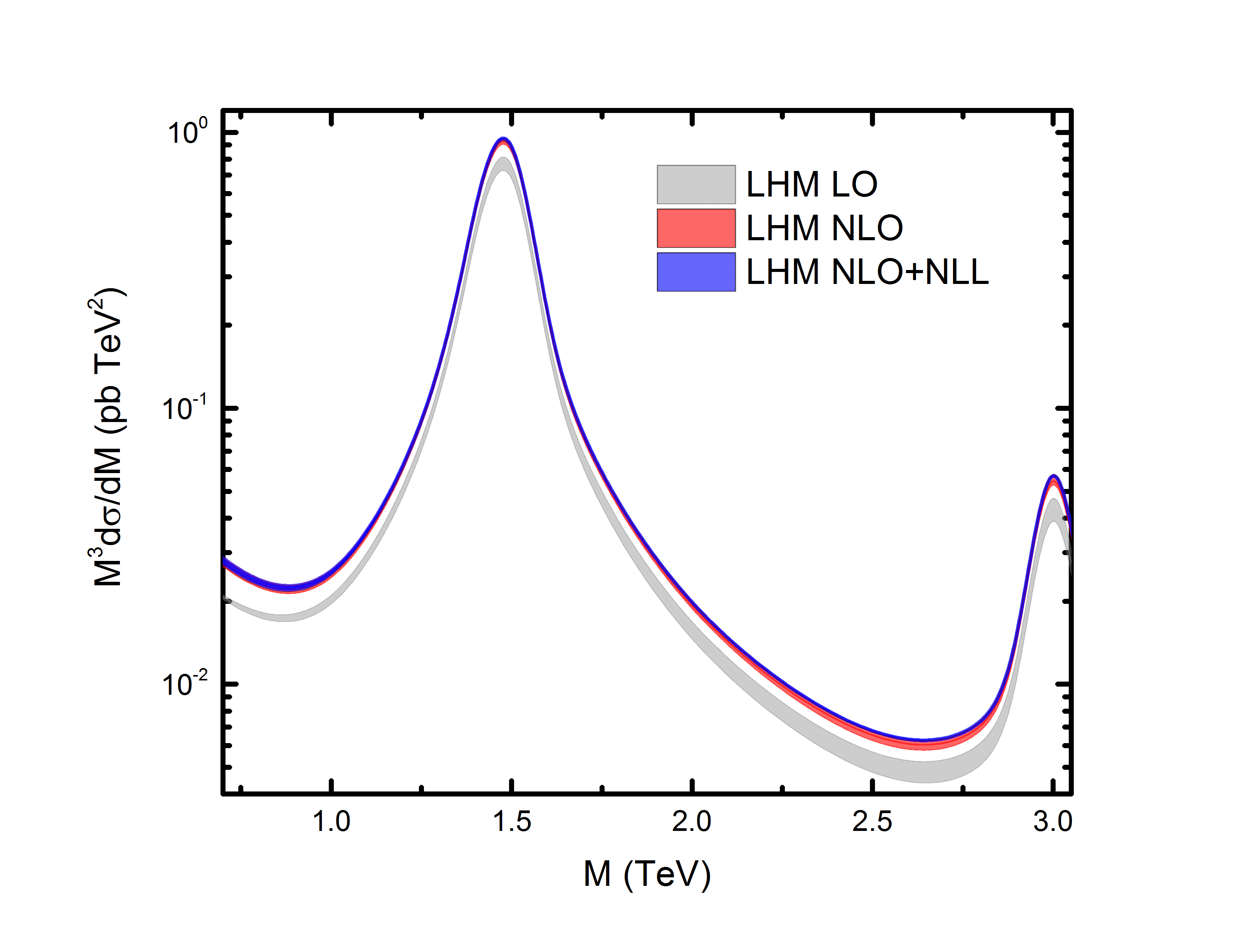}
\caption{\small The LO, QCD NLO, and NLO+NLL corrected $HZ$ invariant mass distributions with the scale varying in the range of $[\frac{1}{2}M,~ 2 M]$ for the $HZ$ production in the LHM at the $14~{\rm TeV}$ LHC. The LO invariant mass distribution range is shown as the gray band, the QCD NLO corrected invariant mass distribution range is shown as the red band, and the QCD NLO+NLL corrected invariant mass distribution range is shown as the blue band.}\label{fig:mmuscale}
\end{center}
\end{figure*}

In Fig.\ref{fig:mmuscale} we depict the $HZ$ invariant mass distributions with the scale uncertainty ranges for the $HZ$ production in the LHM at the $14~{\rm TeV}$ LHC, where we define the scale uncertainty range of the differential cross section of $HZ$ invariant mass by $\mu$ varying in the range of $\mu \in [\frac{1}{2}M,~ 2 M]$. In the figure the LO distribution is drawn as the gray band, the NLO distribution as the red band, and the NLO+NLL corrected distribution as the blue band. Each of the $HZ$ invariant mass distribution bands exhibits two peaks at the positions around $M \sim 1500~{\rm GeV}$ and $M \sim 3000~{\rm GeV}$, respectively. Those peaks come from the diagrams for the \ppHZ process in the LHM that involve exchange of the $A_H$  and $Z_H$ boson, separately. The uncertainty of the LO distribution is evidently the largest as expected, and the uncertainty of the NLO+NLL corrected results is reduced visibly compared with the NLO distribution, especially in the large $HZ$ invariant mass region. From this respect, we can conclude that in studying the $HZ$ invariant mass distribution for the \ppHZ process, the NLO+NLL corrected prediction is more reliable than both the LO and the NLO corrected ones.
%%%%%%%%%%%%%%%%%%%%%%%%%%%%%%%%%%%%%%%%%%%%%%%%%%%%%%%%%%%%

\vskip 5mm
\section{SUMMARY}
\par
In this paper, we calculate the QCD NLO+NLL effects on the $HZ$ production in the LHM at the $14~{\rm TeV}$ LHC including the contribution from the one-loop-induced $gg$-fusion channel. We provide the total cross sections, the transverse momentum, and invariant mass distributions for $HZ$ associated production by combining the QCD NLO corrections obtained by means of perturbative QCD with the resummation of the large logarithmic contributions arising in the small $p_T$ area and the region close to the production threshold. We estimate the theoretical errors for the predictions of the total cross section and kinematic distributions, and find that the QCD NLO+NLL correction improves the scale uncertainties of the LO and pure QCD NLO corrected results. Therefore, we believe that the QCD NLO+NLL corrected predictions are more reliable than the LO and NLO ones. We also show the deviations between the LHM and the SM predictions by providing the transverse momentum and invariant mass distributions in both models up to the QCD NLO+NLL precision. We see from the distributions that the QCD NLO+NLL correction obviously suppresses the relative deviation between the LHM and the SM predictions in the $HZ$ production process, but the LHM signature at the QCD NLO+NLL accuracy would be still observable in precision searches.

\vskip 5mm
\section*{ACKNOWLEDGEMENTS}
This work was supported in part by the National Natural Science Foundation of China (Grants No.11275190, No.11375171, No.11405173, No.11535002).

\vskip 5mm
%\begin{center} {\bf Appendix} \end{center}
\section*{APPENDIX: RELATED LHM COUPLINGS}
\par
The Feynman rules for the coupling vertices in unitary gauge within the LHM related to our work are presented in this appendix. The couplings of the neutral gauge bosons to quarks are expressed in the form as $i\gamma_{\mu}(g_L P_L+g_R P_R)$ where $P_{L,R}\equiv \frac{1}{2}(1\mp \gamma_5)$. The explicit expressions for $g_L$ and $g_R$ are given below.
\begin{equation}\label{Z-coupling-1}
g_{L}^{Z\bar U U}=-\frac{e}{2S_{W}C_{W}}\left
\{1-\frac{4}{3}S_{W}^{2} +\frac{v^{2}}{f^{2}}\left[\frac{c^{2}}{2}
(c^{2}-s^{2})-\frac{5}{2}(c^{\prime2}-s^{\prime2})\left(\frac{8}{15}-\frac{1}{3}c'^{2}\right)
\right]\right\},
\end{equation}
\begin{equation}\label{Z-coupling-2}
g_{R}^{Z\bar U
U}=-\frac{e}{2S_{W}C_{W}}\left\{-\frac{4}{3}S_{W}^{2}-\frac{v^{2}}
{f^{2}}\left[\frac{5}{2}(c^{\prime2}-s^{\prime2})
\left(\frac{2}{15}+\frac{2}{3}c'^{2}\right)\right]\right\},
\end{equation}
\begin{equation}\label{Z-coupling-3}
g_{L}^{Z\bar D
D}=-\frac{e}{2S_{W}C_{W}}\left\{-1+\frac{2}{3}S_{W}^{2}-\frac{v^{2}}{f^{2}}
\left[\frac{c^{2}}{2}
\left(c^{2}-s^{2}\right)+\frac{5}{2}\left(c^{\prime2}-s^{\prime2}\right)\left(-\frac{2}{15}
+\frac{1}{3}c^{\prime2}\right)\right]\right\},
\end{equation}
\begin{equation}\label{Z-coupling-4}
g_{R}^{Z\bar D D}=-\frac{e}{2S_{W}C_{W}}\left\{\frac{2}{3}S_{W}^{2}
-\frac{v^{2}}{f^{2}}\left[\frac{5}{2}(c^{\prime2}-s^{\prime2})
\left(\frac{4}{15}-\frac{2}{3}c'^{2}\right)\right]\right\},
\end{equation}
\begin{eqnarray}\label{ZH-coupling-1}
g_{L}^{A_{H}\bar U
U}=\frac{e}{2s^{\prime}c^{\prime}C_{W}}\left(\frac{2}{15}-\frac{1}{3}c'^{2}\right),\hspace{0.5cm}
g_{R}^{A_{H}\bar U
U}=\frac{e}{2s^{\prime}c^{\prime}C_{W}}\left(\frac{8}{15}-\frac{8}{6}c^{\prime2}\right),
\end{eqnarray}
\begin{eqnarray}\label{ZH-coupling-2}
g_{L}^{A_{H}\bar D
D}=\frac{e}{2s^{\prime}c^{\prime}C_{W}}\left(\frac{2}{15}-\frac{2}{6}c^{\prime2}\right),\hspace{0.5cm}
g_{R}^{A_{H}\bar D
D}=\frac{e}{2s^{\prime}c^{\prime}C_{W}}\left(-\frac{4}{15}+\frac{4}{6}c^{\prime2}\right),
\end{eqnarray}
\begin{eqnarray}\label{ZH-coupling-3}
g_{L}^{Z_{H}\bar U U}=\frac{ec}{2s S_{W}},\hspace{0.5cm}
g_{R}^{Z_{H}\bar U U}=0,\hspace{0.5cm} g_{L}^{Z_{H}\bar D
D}=-\frac{ec}{2s S_{W}}, \hspace{0.5cm} g_{R}^{Z_{H}\bar D D}=0,
\end{eqnarray}
\begin{align}
g^{Z\bar{t}t}_L& =-\frac{e}{2S_{W}C_{W}}\left(1-\frac{4}{3}S_{W}^{2}\right) \,,    &
g^{Z\bar{t}t}_R& =-\frac{e}{2S_{W}C_{W}}\left(-\frac{4}{3}S_{W}^{2} \right) \,,   \\
g^{A_H\bar{t}t}_L&=\frac{e}{2s^{'}c^{'}C_W}\left( \frac{2}{15} - \frac{1}{3}c^{'2}  \right)\,,    &
g^{A_H\bar{t}t}_R&=\frac{2}{2s^{'}c^{'}C_W}\left(  \frac{8}{15}-\frac{4}{3}c^{'2}-\frac{2}{5}\frac{R^2}{1+R^2}  \right)  \,,  \\
g^{Z_H\bar{t}t}_L&=\frac{ec}{2sS_W}\, , & g^{Z_H\bar{t}t}_R&=0  \,,\\
g^{Z\bar{T}t}_L&=i\frac{v}{f}\frac{e}{2S_WC_W}\frac{R^2}{1+R^2}\, ,&  g^{Z\bar{T}t}_R&=0 \,, \\
g^{Z_L\bar{T}T}_L & = \frac{2eS_W}{3C_W} \, , & g^{Z_L\bar{T}T}_R&= \frac{2eS_W}{3C_W}  \, , \\
g^{A_H\bar{T}T}_L & = \frac{e}{2s^{'}c^{'}C_W}\left( \frac{2}{15} -\frac{4}{3}c^{'2} \right)  \,, & g^{A_H\bar{T}T}_R & = \frac{e}{2s^{'}c^{'}C_W}  \left( \frac{2}{15} -\frac{4}{3}c^{'2}+\frac{2}{5}\frac{R^2}{(1+R^2)} \right)  \,,  \\
g^{Z_H\bar{T}T}_L & = {\cal O}(v^2/f^2)   \,, &  g^{Z_H\bar{T}T}_R & = {\cal O}({v^2/f^2}) \,.
\end{align}

\par
The couplings between the Higgs boson and quarks are expressed as
\begin{align}
g^{H\bar{t}t}&=-i\frac{M_t}{v} \left[  1-\frac{s_0^2}{2} +\frac{v}{f}\frac{s_0}{\sqrt{2}} -\frac{2}{3}\frac{v^2}{f^2}
+\frac{v^2}{f^2}\frac{R^2}{1 + R^2} \left(1 + \frac{R^2}{1+ R^2} \right )
 \right]  \, ,  \\
 g^{H\bar{T}t}&=\frac{M_t}{v}\frac{v}{f}\left( 1 + \frac{R^2}{1 + R^2} \right) P_R + \left(\frac{M_t}{v}R\right)P_L \,,  \\
g^{H\bar{t}T}&=-\frac{M_t}{v}\frac{v}{f}\left( 1 + \frac{R^2}{1 + R^2} \right) P_L - \left(\frac{M_t}{v}R\right)P_R \,, \\
 g^{H\bar{T}T} & = -i\frac{M_t}{v}\frac{v}{f}R \left(  1 + \frac{R^2}{1+R^2} \right) \,,
\end{align}
where $s_0=\frac{\sqrt{2}}{2}\frac{v}{f}\chi$, $R$ is input parameters introduced in the LHM, and the mass of the extra top quark partner is expressed as $M_T=\frac{M_tf}{v}\frac{1+R^2}{R}$. $U$ and $D$ represent the up-type $(U=u,c,t)$ and down-type $(D=d,s,b)$ quarks, respectively. The couplings between neutral gauge boson and Higgs boson are expressed as
\begin{equation}
g^{HZZ}=\frac{ie^{2}v g_{\mu\nu}}{2S_{W}^{2}C_{W}^{2}}
\left\{1-\frac{v^{2}}{f^{2}}\left[\frac{1}{3}-\frac{3}{4}\chi^{2}+\frac{1}{2}(c^{2}-s^{2})^{2}+
\frac{5}{2}(c^{\prime2}-s^{\prime2})^{2}\right]\right\},
\end{equation}
\begin{equation}\label{Z_H-A_H-ZH}
g^{HZA_{H}}=-\frac{ie^{2}v
g_{\mu\nu}}{2S_{W}C_{W}^2}\frac{c^{\prime2}-s^{\prime2}}{2s^{\prime}c^{\prime}},
\hspace{0.8cm} g^{HZZ_{H}}=-\frac{ie^{2}v
g_{\mu\nu}}{2S_{W}^2C_{W}}\frac{c^{2}-s^{2}}{2sc}.
\end{equation}

\par
The partial decay widths for $V_{H} \to f\bar{f}$ and $V_{H} \to ZH$ ($V_{H}~(V_H=Z_H,A_H)$) can be expressed as \cite{width}
\begin{equation} \label{VH-width-1}
\Gamma(V_{H}\to f\bar f)=\frac{N_{c}}{12\pi}
\left[(g^{V_{H}\bar f f}_{v})^{2}(1+2r_{f})+(g^{V_{H}\bar f f}_{a})^2(1-4r_{f})\right]\sqrt{1-4r_{f}}M_{V_{H}},
\end{equation}
\begin{equation}\label{VH-width-2}
\Gamma(V_{H}\to ZH)=\frac{(g^{V_H})^2}{192\pi}\sqrt{\lambda}
\left[(1+r_{Z}-r_{H})^{2}+8r_{Z}\right]M_{V_{H}},
\end{equation}
where $N_c = 3$ is the color factor, $g^{V_Hf \bar f}_v=(g^{V_Hf \bar f}_R+g^{V_Hf \bar f}_L)/2$,
$g^{V_Hf \bar f}_a=(g^{V_Hf \bar f}_R-g^{V_Hf \bar f}_L)/2$,
$g^{A_H}=g^{\prime}(c^{\prime 2}-s^{\prime 2})/(2c^{\prime}s^{\prime})$,
$g^{Z_H}=g(c^{2}-s^{2})/(2cs)$, $\lambda=1+r_{Z}^2+r_{H}^2-2r_{Z}-2r_{H}-2r_{Z}r_{H}$,
and $r_{i}=X_{i}^2/M^2_{V_{H}}$ $(X_{i}=m_f,M_Z,M_H)$.
Since in our investigated parameter space the $V_H \to T\overline{T}$ and $V_H
\to \overline{T}t(T\bar{t})$ decays are kinematically forbidden, we
assume that the total decay width $\Gamma_{V_H}~(V_H=Z_H,A_H)$ is
the sum of $\Gamma(V_{H}\to f\bar f)$ and $\Gamma(V_{H}\to ZH)$,
where $f=u,d,c,s,b,t,$ $e,\mu,\tau,\nu_e,\nu_{\mu},\nu_{\tau}$.

\end{document}